\documentclass[twocolumn]{aastex631}

\newcommand{\chandra}{{\it Chandra}}

\newcommand{\xmm}{{\it XMM}-Newton}

\newcommand{\nustar}{{\it NuSTAR}}

\newcommand{\nova}{V1674~Her} 

\begin{document}
\title{Nova V1674 Her as a fast magnetic rotator}

\correspondingauthor{G. J. M. Luna}
\email{juan.luna@unahur.edu.ar}

\author[0000-0002-2647-4373]{G. J. M. Luna}
\affiliation{Universidad Nacional de Hurlingham (UNAHUR). Laboratorio de Investigación y Desarrollo Experimental en Computación, Grupo de Astrofísica, Av. Gdor. Vergara 2222, Villa Tesei, Buenos Aires, Argentina}
\affiliation{Consejo Nacional de Investigaciones Científicas y Técnicas (CONICET), Argentina.}

\author[0000-0003-1563-9803]{M. Orio}
\affiliation{INAF-Osservatorio Astronomico di Padova, vicolo Osservatorio, 5,
35122 Padova, Italy }
\affiliation{Department of Astronomy, University of Wisconsin, 475 N. Charter Str., Madison WI 53706, WI, USA}

\author[0009-0004-3869-2425]{A. Dobrotka}
\affiliation{Advanced Technologies Research Institute, Faculty of Materials Science and Technology in Trnava, Slovak University of Technology in Bratislava, Bottova 25, 917 24 Trnava, Slovakia}

\author{J. Magdolen}
\affiliation{Advanced Technologies Research Institute, Faculty of Materials Science and Technology in Trnava, Slovak University of Technology in Bratislava, Bottova 25, 917 24 Trnava, Slovakia}

\begin{abstract}
We report on the results of our \xmm\ and \nustar\ campaign targeting \nova\ in October 2023. These observations were timed to investigate the post-nova quiescent state 28 months after the 2021 outburst, although the optical luminosity was still slowly declining. The X-ray spectra are well-modeled by a multi-temperature thermal plasma. While in the best fit temperature we find that the highest plasma temperature is indicative of a massive white dwarf (\(M_{\text{WD}} \approx 1.35 M_\odot\)), we caution that there is a degeneracy between the plasma temperature and the metallicity in the spectral fit, hindering a definitive constraint.  Alternatively, the continuum can be phenomenologically described by a Comptonized component, although this model neglects the thermal bremsstrahlung component.
Timing analysis reveals a stable 501.31 $\pm$ 0.19 s modulation in the post-quiescent phase, whose amplitude diminishes with increasing energy; this is quite typical in intermediate polars (IP). The phase alignment between the quiescent and early post-outburst \chandra\ spin-folded light curves suggests that the soft X-ray emission originates from the same polar regions, which quickly re-established accretion following the eruption. \nova\ belongs to a growing class of novae that are also IPs, and is one of four in which the modulation with rotation was observed during the outburst supersoft X-ray phase. We propose that the fast rise in outburst was due to a fast or centrifugal magnetic rotator wind, causing rapid mass loss instead of the commonly assumed radiation pressure wind.
\end{abstract}



\section{Introduction} \label{sec:intro}

The outburst of \nova\ was reported  by Seiji Ueda (Kushiro, Hokkaido, Japan) on 2021 June 12.537 UT, when it reached a magnitude V=6.2 within only few hours \citep[see][and AAVSO light curve]{Quimby2024}. The nova distance has not been well determined yet: the GAIA distance is 6.00$_{-2.76}^{+3.8}$ kpc. Because of strong neon lines in the optical spectra, \citet{Wagner2021} suggested that the nova occurred on an oxygen-neon white dwarf (WD), which should have a mass larger than $\simeq$1.1 M$_\odot$. The high supersoft source effective temperature suggests a very massive WD \citep{Drake2022}; \citet{2024MNRAS.527.1405H} obtained  a mass of even 1.36 M$_\odot$ by modeling the optical spectrum in outburst with the CLOUDY photoionization code. \citet{Hachisu2026V1674her} suggested a mass of 1.35 M$_\odot$.

The orbital period is 0.15302(2) days \citep{2021ATel14835....1S, 2021ATel14856....1P,Luna2025tess}. A modulation with a period of 503.9 seconds detected in X-rays immediately after the outburst, in close agreement with the 501.42 s periodicity reported by \citet{2021ATel14720....1M} in pre-outburst $ZTF$ data, has been attributed to the WD spin. The detection of coherent modulations during accretion is typical of an intermediate polar (IP), in which the white dwarf has a magnetic field of at least a few 10$^5$ Gauss and the accretion disk is disrupted, with accretion streamed to the poles. On the other hand, periodicities of several minutes during the supersoft X-ray phase - while the nuclear burning under the surface continues for some time after a thermonuclear runaway- have been detected in several other novae \citep{Orio2022} and three of them are now confirmed IPs \citep[V4743~Sgr,V2491~Cyg,V407~Lup;][]{Zemko2016,Dobrotka2017,Zemko2015,Aydi2018,Orio2024}. In the observations of the novae listed in \citet{Orio2022} with periods of tens of seconds observed during outburst, or in the more recent V1716 Sco \citep{Worley2025} periods of tens of seconds have not been detected yet in quiescence. The recent detection of a 37.691\,s modulation in the quiescent optical light curve of YZ~Ret constitutes an interesting case that warrants additional investigation \citep{Luna2026YZRET}.

We present results of the spectral and timing analysis of X-ray observations of \nova \ obtained with \xmm\ + \nustar\ after it returned to the quiescent state. In Section \ref{sec:data} we present the observations and how the spectra and light curves were extracted, while Section \ref{sec:res} highlights the results of their analysis. In Section \ref{sec:disc} we discuss the results in the context of the aftermath of novae in magnetic cataclysmic variables. We present conclusions in Section \ref{sec:conc}.

\begin{figure*}
\begin{center}
\includegraphics[scale=0.4]{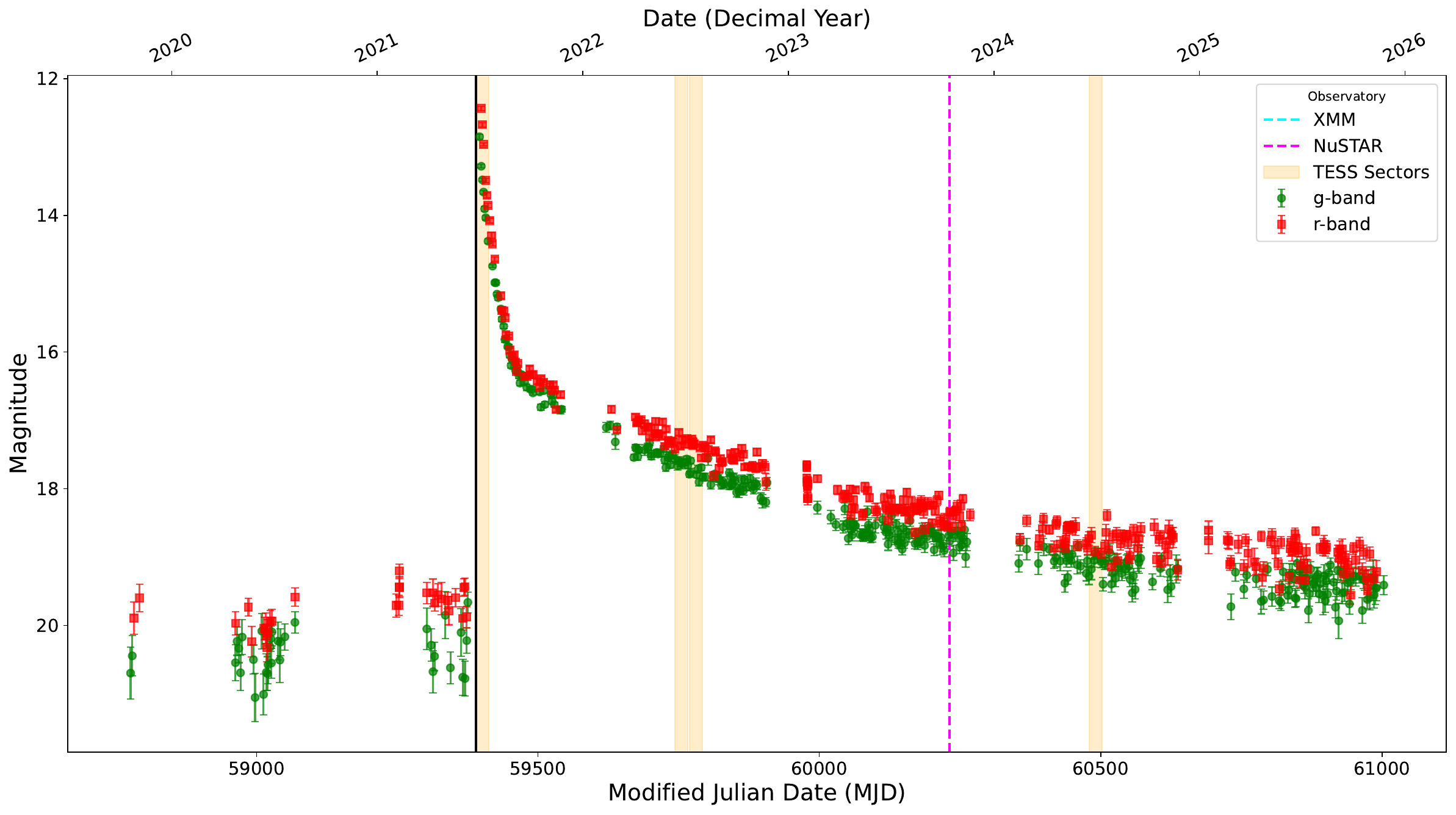}
\caption{Historical $ZTF$ light curve of \nova\ including its 2021 nova eruption (solid vertical line). The epoch of the \nustar\ and \xmm\ observations are marked with the vertical dashed lines; since both observations were performed quasi-simultaneously, the two lines overlap and are indistinguishable in the figure. The shaded areas mark the dates of the $TESS$ Sectors observations discussed in \citet{Luna2025tess}.}
\label{fig:ztf}
\end{center}
\end{figure*}

\section{Observations and data analysis} \label{sec:data}
Figure \ref{fig:ztf} shows the epoch at which the observations presented and analyzed here were taken in the context of the outburst evolution and of the nova returning to quiescence as followed by the Zwicky Transient Facility \citep[ZTF;][]{Graham2019}. We also show the dates of the $TESS$ observations analyzed in \citet{Luna2025tess}. Our \xmm\ + \nustar\ observations were performed during the last portion of the declining phase, after mass loss ceased but the \nova\ was still nearly two magnitudes above the pre-outburst optical flux level as measured in  the $ZTF$. 

\subsection{\xmm}

We observed \nova\ with all cameras on board the \xmm\ satellite on October 14, 2023, UT 00:39:36 during 56 ks (ObsID: 0931400101, PI: M. Orio). The EPIC cameras were operated in Full Frame mode with the thin filter while the optical monitor (OM) was operated in image mode with the $UVM2$ filter. We selected standard grades for the $pn$ (0–4) and $mos$ 1/2 (0–12) cameras. The RGS1 and RGS2 cameras were operated in spectroscopy mode but rendered spectra with low signal-to-noise  and are not useful for spectral analysis. Broad band X-ray spectra of the source and of the background were extracted in the EPIC cameras from circular regions with 20$^{\prime\prime}$ and 45$^{\prime\prime}$ radii, respectively. The source region was centered on \nova\ SIMBAD coordinates ($\alpha$(J2000)=18h 57m 30.98s, $\delta$(J2000)=+16$^{\circ}$ 53$^{\prime}$ 39.5$^{\prime\prime}$), while the background was selected in a nearby, source-free area. We used the \texttt{rmfgen} and \texttt{arfgen} scripts to build the redistribution matrices and ancillary EPIC responses. The energy channels were grouped in such a way that each spectrum had at least 25 counts per bin. The barycentric correction was applied to all EPIC event arrival times using the SAS task \texttt{barycen}, ensuring all timings are reported in Barycentric Dynamical Time.

\subsection{\nustar}
\nustar\ observed \nova\ quase-simultaneously with \xmm\ during 41.8 ks on October 14 2023 UT04:16:09 using both the FPMA and FPMB modules (ObsID: 30901021002, PI: M. Orio). We extracted spectra and baricenter-corrected light curves from the source and the background using the \texttt{nuproducts} script. Source photons were selected from a circular region with 60$^{\prime\prime}$ radii centered on the same SIMBAD coordinates as the region used to extract \xmm\ products. Source and background light curves were binned every 10 seconds. 
\section{Results} \label{sec:res}
\subsection{Spectral analysis}
Figure 2 shows the spectrum measured with the {\sl XMM-Newton} EPIC cameras (0.3-10 keV range) and with the NuSTAR cameras in the 3-40 keV range (no source photons above the background were detected at energies higher than 40 keV) and the fit with Model A whose parameters are listed in Table 1. The parameters of the spectral fits were drawn from modeling the combined \xmm\ and \nustar\ data with two competing models: Model A includes a blackbody and a multi-temperature optically thin thermal plasma in an isobaric cooling flow \citep[\texttt{mkcflow} in XSPEC][]{mkcflow1988}. The second model, Model B, has been recently proposed by \citet{2021ApJ...911...80M}. These authors explored the possibility that the continuum X-ray emission of IPs is described by a Comptonization model, \texttt{compTT} in XSPEC. In this model seed photons are scattered off by the electrons of an optically thick plasma. The seed photons are provided by the soft X-rays/UV emission of the footprint of the accretion column on the white dwarf surface, while the optically thick plasma is in the accretion column. In both cases, in addition to the absorbing column density modeled with \texttt{tbabs} in XSPEC, we also used a partially covering absorber in the fit, due to the accretion column/curtain (\texttt{pcfabs}). The covering fraction turned out to be of the order of 60\% in both models.

Recently, \citet{Olbemo2026} modeled the  \xmm\ + \nustar\ data obtained by us and already included in the {\it HEASARC} archive, using the \texttt{ipolar30kK} models' grid calculated by \citet{Suleimanov2025}. 
Our own fit using the same model yielded different parameters than those listed in their Table 3. We list the parameters we obtained under Model C in Table \ref{tab:1}. In any case, we suggest that the relatively large difference in the physical parameters yielded by our fit with Model A and those derived by \citet{Olbemo2026} are explained by taking into account the lack of prominent iron lines in the spectrum and by the background limited signal-to-noise ratio in the spectrum above 40 keV. The subtle continuum curvature below 40 keV makes it difficult to unambiguously decouple a very high \(kT_{\max}\) from an extremely low metallicity (\(Z/Z_\odot < 0.01\)), which is required to suppress the H- and He-like iron lines. Following \citet{Suleimanov2025}'s, we fit the 15--40 keV {\sl NuSTAR} range, thus excluding the counts in the 40-78 keV range included instead by the above authors. With the \texttt{ipolar30kK} grid, we obtained a value  M$_{WD}$=1.26$_{-0.56}^{+0.11}$\, M$_{\odot}$, which does not constrain the mass significantly; with both this and our \texttt{mkcflow} model, we also could not constrain the temperature, obtaining only a lower limit, kT$_{max}>$ 25 keV. 
The other important difference between our fitting procedures is the \citet{Olbemo2026}'s  reflection component with almost solar abundance, even if the spectrum lacks a detectable fluorescent Fe K$\alpha$ \(6.4\text{ keV}\) line. In order to robustly test the presence or absence of this line, we performed a Monte Carlo simulation tracking the \(\Delta\chi^2\) distribution under the null hypothesis \citep{Protassov2002ApJ...571..545P}. The improvement (\(\Delta\chi^2 \approx -15\)) is entirely consistent with the simulated null distribution, confirming that a \(6.4\text{ keV}\) feature is not significantly detected in the data. The non solar-abundance of iron is also consistent with the extremely low metallicity (\(Z/Z_\odot < 0.01\)) yielded by our \texttt{mkcflow} fit, which would naturally suppress fluorescent lines, regardless of the reflection amplitude.  We also note that the interpretation of \citet{Olbemo2026} of cos($\theta$) as the orbital inclination instead of the observer-reflector angle is technically incorrect, although this choice has a negligible impact on the derived WD mass.
\citeauthor{Suleimanov2025}'s model assumes that the post-shock region cools by bremsstrahlung continuum emission and does not include cooling by line emission \citep[see discussion in][]{Luna2015A&A...578A..15L}. Although this does not significantly affect the white dwarf mass value yielded by the model, the X-ray luminosity and consequently the mass accretion rate are underestimated, the latter by about 20-30\% \citep[see discussion in][]{Suleimanov2019}.

\begin{figure}
\begin{center}
\includegraphics[scale=0.32]{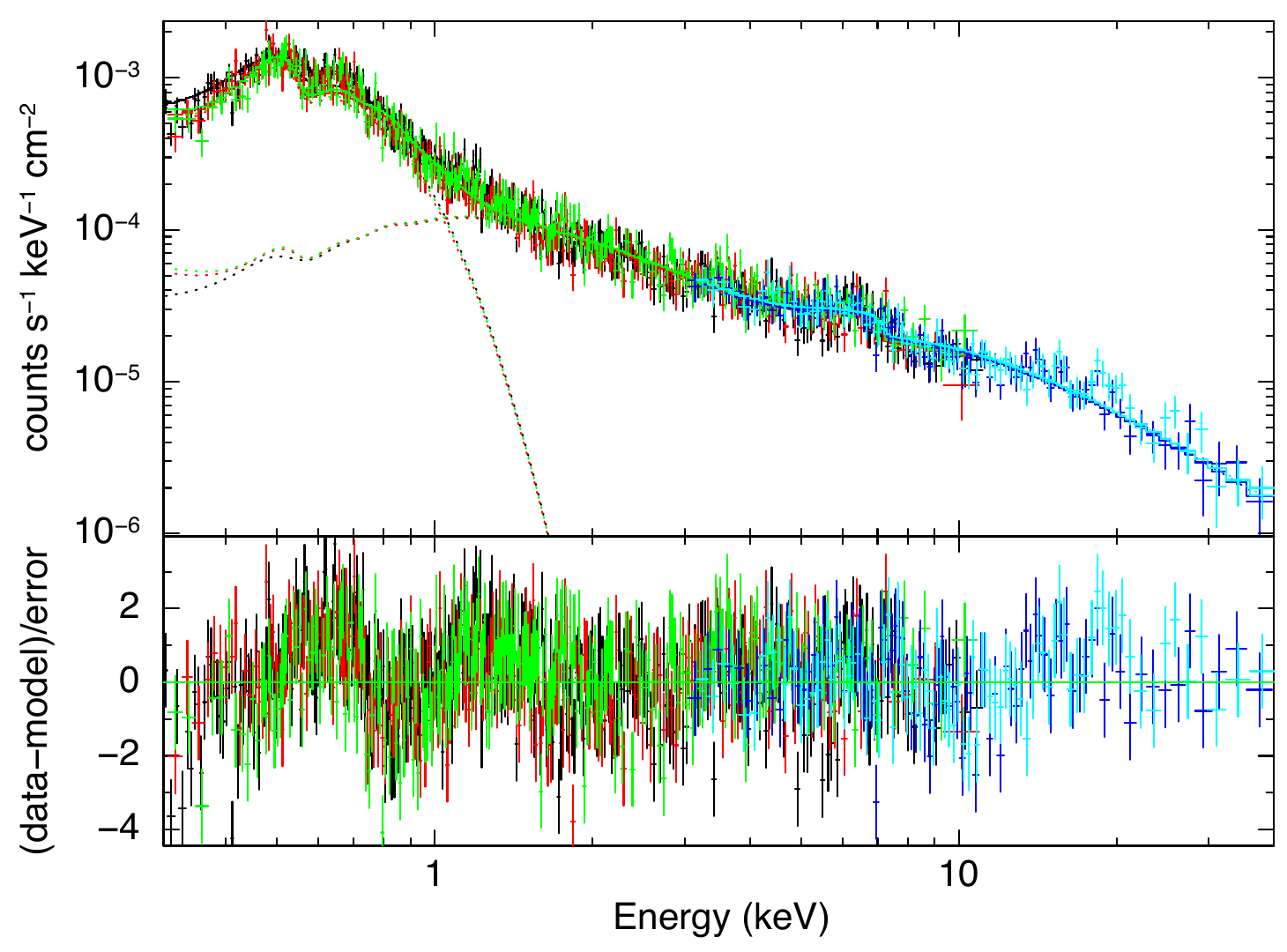}
\caption{\xmm\ (pn, mos1 and mos2) plus \nustar\ FPMA and FPMB spectra. The dotted lines show the contribution of each model component (blackbody and cooling flow) to the total spectrum. }
\label{fig:spec}
\end{center}
\end{figure}

\renewcommand{\arraystretch}{1.3}
\begin{deluxetable*}{lccc}
\tablecaption{ \label{tab:1} Parameters of the fits to the \xmm\ + \nustar\ spectra. The unabsorbed X-ray flux and luminosity for the blackbody, the cooling flow and the comptonize flux are calculated in the 0.3-40 keV energy band. Elemental abundances are relative to the solar values tabulated by \citet{2000ApJ...542..914W}. Luminosity and $\dot{m}$ are determined assuming a distance of 6 kpc, the GAIA distance and also average of the values published \citep[see e.g.][and references therein]{Aydi2025,2023MNRAS.521.5453S}. Statistical errors are calculated at the 90\% confidence level. }
\tablehead{
\colhead{}&  Model A\tablenotemark{a} & Model B\tablenotemark{b} & Model C\tablenotemark{c}}
\startdata
\hline
N$_{H}$ (F.C.) (10$^{22}$ cm$^{-2}$)  & 0.31$\pm$0.01 & 0.32$\pm$0.02 & 0.26$\pm$0.02\\
N$_{H}$ (P.C.) (10$^{22}$ cm$^{-2}$)  & 53$\pm$6 &  34$\pm$5 & 63$_{-23}^{+33}$ \\
Covering fraction  & 0.64$\pm$0.02 & 0.58$\pm$0.06 & 0.47$\pm$0.11\\
kT$_{bb}$[keV]   & 0.096$\pm$0.002 & 0.094$\pm$0.003 & 0.13$\pm$0.01 \\
Area$_{bb}$ (km$^2 \times 10^{4}$)  & 8.84$\pm$0.11  & 8.51$\pm$0.09 & 1.4$_{-0.1}^{+0.2}$ \\ 
kT$_{\rm max}$ (keV)   & $\geq$69 & 11$_{-2}^{+9}$ &\nodata\\
Z/Z$_{\odot}$   & $<$0.01 & \nodata &\nodata\\
$\tau$  & \nodata & 7$\pm$2 &\nodata\\
$\dot{m}_{thin}$ [10$^{-9}$ M$_{\odot}$ yr$^{-1}$]   & 3.3$_{-0.2}^{+0.5}$ & \nodata &\nodata\\
Edge energy  [keV]& \nodata & \nodata &0.92 \\
Edge $\tau$ & \nodata & \nodata & 0.37$\pm$0.06\\
Edge energy  [keV]& \nodata & \nodata &0.74\\
Edge $\tau$ & \nodata & \nodata & 0.80$\pm$0.03\\
Reflect Abund.\tablenotemark{d} & \nodata & \nodata & 0.65$\pm$0.02\\
\texttt{ipolar30kK} fall\_height R$_{WD}$ & \nodata & \nodata & $\gtrsim$110\\
\texttt{ipolar30kK} M$_{WD}$ [M$_{\odot}$]& \nodata & \nodata & $\gtrsim$1.15\\
\texttt{ipolar30kK} norm & \nodata & \nodata & $2.5_{-2.0}^{+3.5}\times$10$^{-29}$\\
$\chi^{2}$/d.o.f   & 1.28 & 1.24 &1.64\\
F$_{bb} \times 10^{-11}$ erg cm$^{-2}$ s$^{-1}$   & 1.80$\pm$0.02 & 1.80$\pm$0.03 &0.52$\pm$0.04\\
F$_{hard} \times 10^{-11}$ erg cm$^{-2}$ s$^{-1}$  & 0.76$\pm$0.03 & 0.66$\pm$0.02 &0.48$\pm$0.03\\
L$_{bb} \times 10^{34}$ erg s$^{-1}$ & 7.7$\pm$0.5 & 7.6$\pm$0.5 &2.2$\pm$0.1\\
L$_{hard} \times 10^{34}$ erg s$^{-1}$ & 3.2$\pm$0.1 & 2.8$\pm$0.1  &2.1$\pm$0.1\\
\enddata
\tablenotetext{a}{\texttt{pcfabs$\times$TBabs$\times$(bbody+mkcflow)}}
\tablenotetext{b}{\texttt{pcfabs$\times$TBabs$\times$(compTT+bbody)}}
\tablenotetext{c}{Model fit from \citet{Olbemo2026}, \texttt{constant$\times$pcfabs$\times$edge$\times$edge$\times$TBabs$\times$(bbody + reflect$\times$(gaussian + ipolar30kK))}} 
\tablenotetext{d}{The other parameters of the \texttt{reflect} component are not listed and fixed at their values from \citet{Olbemo2026}.}
\end{deluxetable*}

\subsection{Timing analysis } \label{sec:lc}
\begin{figure}
\includegraphics[scale=0.42]{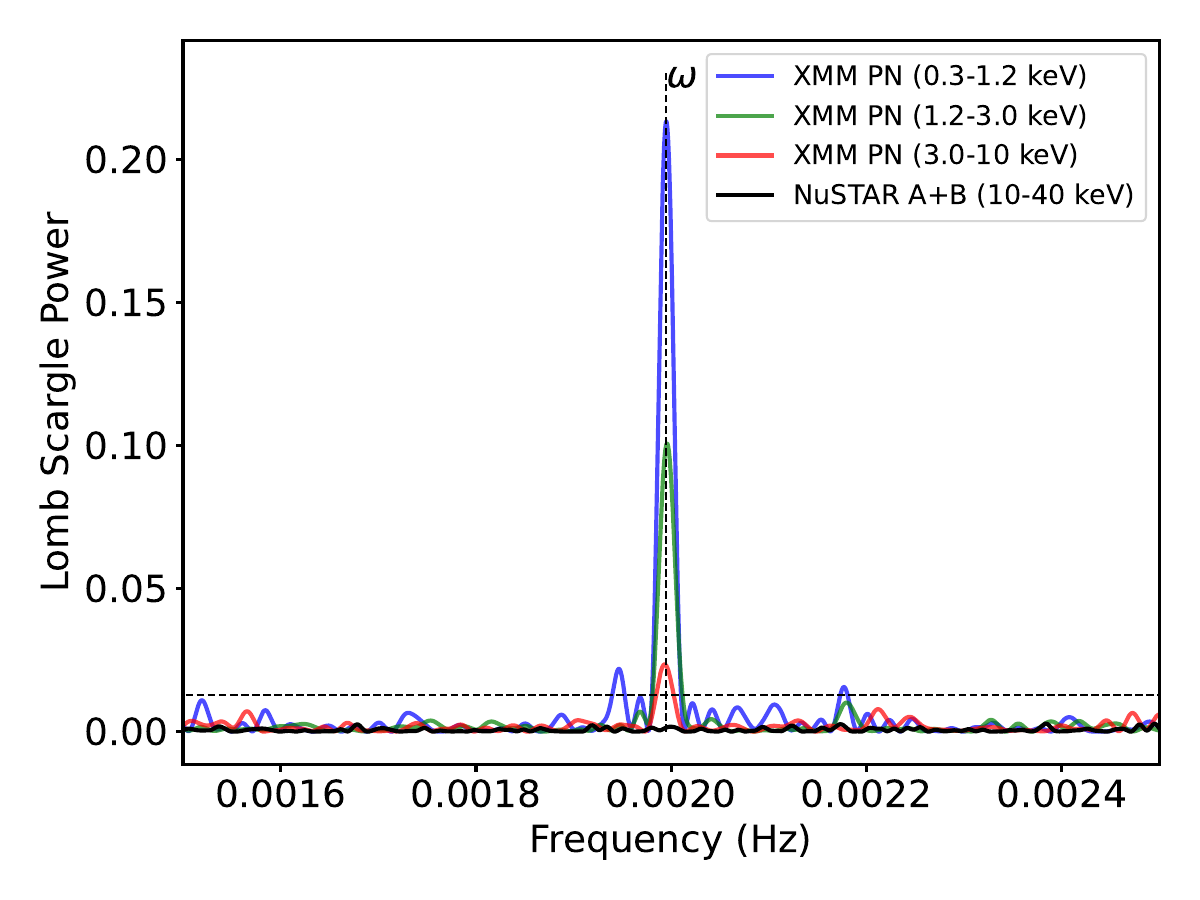}
\caption{Lomb-Scargle Power Spectrum obtained from the light curves of the \xmm\ observation in the 0.3-1.2 keV ($pn$: blue) and 1.2-3 keV ($pn$: red), 3-10 keV ($pn$: green) energy ranges and the \nustar\ FPMA + FPMB (black) in the 10-40 keV energy range. The horizontal dashed line marks the 99.9\% FAP level from the \nustar\ power spectrum while the vertical dashed line marks the frequency corresponding to a period of 501.31\,s, which is identified as the white dwarf spin period. }
\label{fig:psd}
\end{figure}
Fig. 3 shows the results obtained using  the Lomb-Scargle algorithm as implemented in the \texttt{Astropy} \citep{Astropy2022} library to search for periods in the power spectrum. We detected a clear period of 501.31 $\pm$ 0.19 s in the 0.3-1.2 keV energy band. This period was also detected in the 1.2-3 and 3-10 keV energy bands in the \xmm/$pn$ light curve but was not detected in the 10-40 keV energy range in the \nustar\ light curve above the 99.9\% false alarm probability level. \citet{Olbemo2026} report a periodicity in the \nustar\ data that corresponds to their analysis of the full \nustar\ energy range, from 3-79 keV. The \xmm\ light curve is modulated at this period with a pulsed fraction of 60\% in the 0.3-1.2 keV energy range, but the amplitude of the modulation decreases to 58\% in the 1.2-3.0 keV energy range and to 47\% in the 3.0-10 keV energy range as covered with \xmm \footnote{ ($PF = \frac{I_{\text{max}} - I_{\text{min}}}{I_{\text{max}} + I_{\text{min}}}$, where $I$ is the count rate.)}. The decrease of the pulse fraction with energy is consistent with a modulation that is due to the energy-dependent absorption of the post-shock accretion column emission by pre-shock material as observed in other IPs \citep{2017PASP..129f2001M}. 

\begin{figure}
\begin{center}
\includegraphics[scale=0.42]{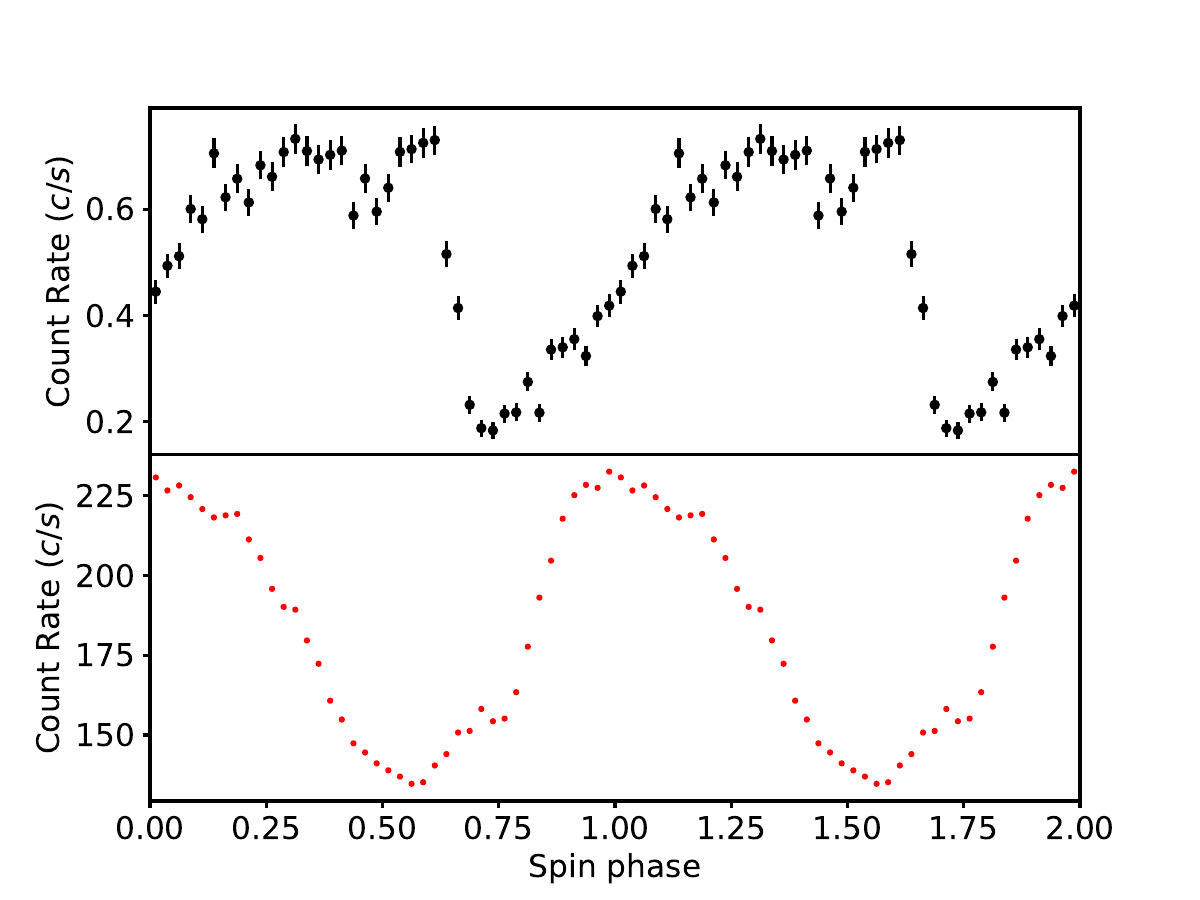}
\caption{{\it Top panel:} X-ray light curve of the EPIC-$pn$ camera in the 0.3-1.2 keV energy range, folded with the 501.31 $\pm$ 0.19\,s period derived with the Lomb-Scargle periodograms. {\it Bottom panel:} The 0 order light curve of the observation done with {\sl Chandra} on day 37 of the outburst, measured with the HRC-S camera, folded at the period of 501.72\,s reported by \citet{Drake2022}. Both light curves were folded using the ephemeris by \citet{2024CoSka..54b.128D}.}
\label{fig:lc}
\end{center}
\end{figure}

To compare the pulse profiles across epochs, we folded both the 2021 \chandra\ (day 37) and 2024 \xmm\ light curves using the same spin ephemeris given by \citet{2024CoSka..54b.128D}. The result is shown in Fig.~\ref{fig:lc}. Despite the transition from a nuclear-burning SSS phase to an accretion-powered quiescent state, the pulse minima show a remarkably small phase shift. This stability indicates that the magnetic poles remained the primary site of both nuclear burning and subsequent accretion, with only a small migration of the 'hot spots' over a 2.3-year baseline. This provides strong observational support for the hypothesis that the large-scale magnetosphere remained largely intact and acted as a rigid anchor for the accretion flow throughout the eruption

\begin{figure}
\includegraphics[scale=0.5]{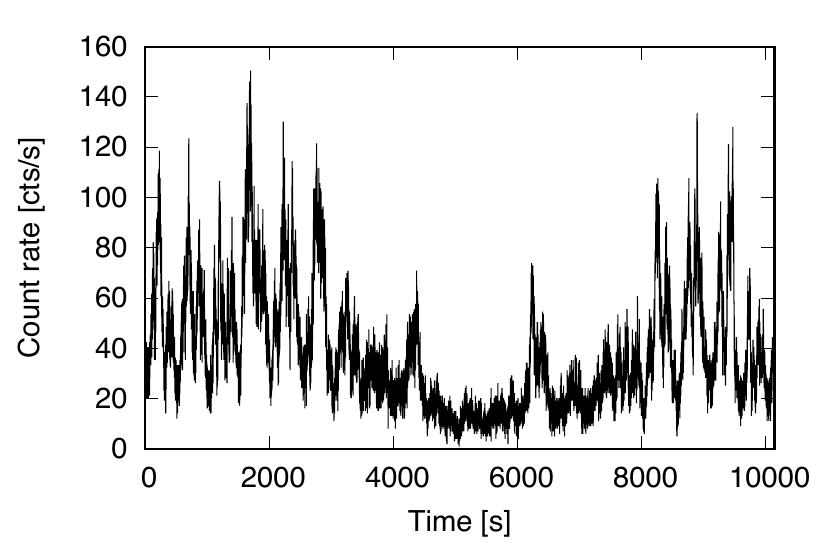}
\caption{\chandra\ light curve of day 28 of the outburst, extracted from the 0 order of the HRC-S detector with a 1\,s bin size.}
\label{lc_day28}
\end{figure}

\begin{figure}
\includegraphics[scale=0.5]{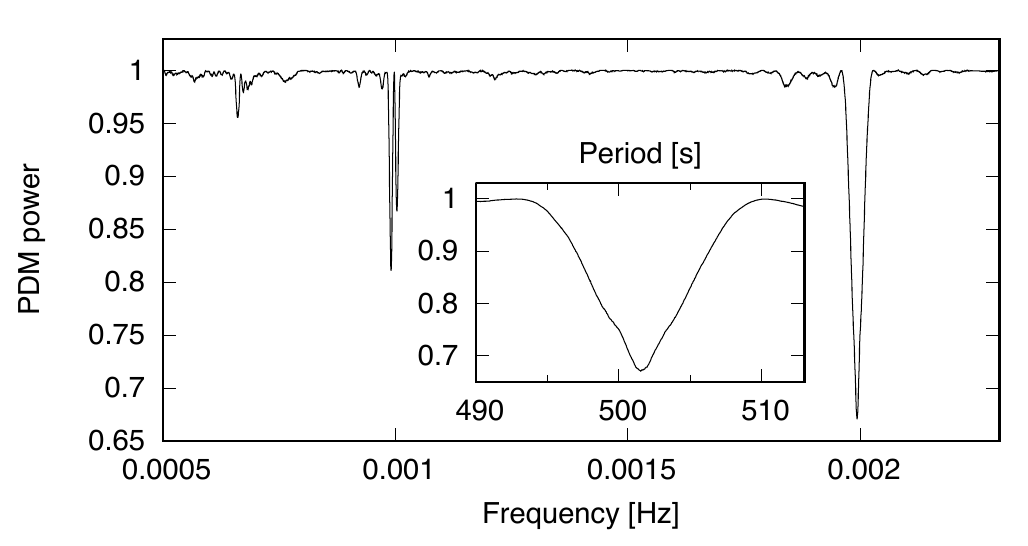}
\caption{PDM power of the \chandra\ light curve of day 28.}
\label{periodogram_pdm}
\end{figure}

\citet{Drake2022} analyzed the \chandra\ outburst light curve of day 37 after the optical maximum, finding a  modulation with a $501.72 \pm 0.11$ period. The authors also mention a \chandra\ light curve of day 28 of the outburst having a modulation with a 503.9\,s period \citep{Maccarone2021}. This value is clearly different from the 501.31\,s or 501.72\,s periods derived in this paper and by \citet{Drake2022}, respectively. Since the light curve is very complex and shows variable amplitude (as clearly seen in Fig.~\ref{lc_day28}),  we dedicate an additional, in-depth analysis to it.
The Lomb-Scargle algorithm yields a period of 506.73\,s; however, this is based on simple sine-wave variability, and does not take into account a more complicated shape with more harmonics. In order to account for the underlying irregular variability in the light curve and also for the varying amplitude of the modulation (see Fig. \ref{lc_day28}), as a second step we fitted it with a model based on the sum of two sine functions with periods $p$ and $p/2$, where the amplitudes are described by polynomials\footnote{Empirically we used polynomials of the 5-th degree.} (P$_1$ and P$_2$):
\begin{equation}
\psi = {\rm P}_0 + {\rm P}_1\,{\rm sin}(2\,\pi\,t/p + \phi_1) + {\rm P}_2\,{\rm sin}(2\,\pi\,t\,2/p + \phi_2),
\label{lc_model}
\end{equation}
and $\psi$ is the flux, P$_0$ is the long-term trend and $\phi$ are the phases. We obtained a 503.8\,s period, close to the one derived by \citet{Maccarone2021}. Using the same model with only one sine function, a polynomial amplitude P$_1$ and a fundamental period $p$ we derived a 505.89\,s value, which is very similar to the value derived from the Lomb-Scargle algorithm. Clearly, the higher harmonics must be taken into account. Finally, when we used a phase dispersion minimization method (PDM) by \citet{stellingwerf1978}, which is more suitable for non-sinusoidal time variations, the resulting periodogram - depicted on Fig.~\ref{periodogram_pdm} - shows a minimum peak yielding a period of 501.53\,s. We conclude that this is the reliable value for this observation, consistent with the periods of 501.31\,s and 501.72\,s derived in this paper and by \citet{Drake2022}, respectively.
We estimated the uncertainty as the standard deviation of the periods measured in 1,000 simulated light curves. We fitted the data shown in Fig.~\ref{lc_day28} with a multi-component model as in Equation~\ref{lc_model} and we generated 1,000 simulations with different Poisson noise. We obtained values of 0.10 and 0.21 using Lomb-Scargle and PDM, respectively. Since the periodicity itself is measured with the PDM, we assumed that xx is the uncertainty estimate. However, the light curve of day 28 is very variable; even the multi-component model describes the data only approximately. The mean value of the periodicity after 1,000 simulations is 501.94 and 500.77\,s with Lomb-Scargle and PDM, respectively even if a value 501.53\, was used. Apparently, even the simple multi-component model we used deforms the modulations and both method yields slightly different method like in the case of observed data, Therefore, the uncertainty may still be underestimated and a very complex model may be required to account for such complex amplitude variability.
\section{Discussion} \label{sec:disc}
\subsection{The Physical Nature of the X-ray Emission: Thermal Shock vs. Comptonized Column}
Our spectral analysis of the late-time X-ray emission from \nova\ reveals two statistically comparable but physically distinct interpretations. In the standard framework for Intermediate Polars (IPs), the X-ray emission originates from a multi-temperature post-shock region as the accreted material settles onto the white dwarf (WD) surface (Model A). The high maximum temperature derived ($kT_{max} \geq 69$ keV) is consistent with a shock forming above a massive white dwarf ($M_{WD} \approx 1.35 M_\odot$). The observed suppression of the Fe K$\alpha$ complex (Fig. \ref{fig:spec}) in this model is interpreted as a result of the extreme temperatures and high ionization state of the plasma, where the emissivity of the iron lines is naturally lower. This model aligns with the ``fast" nature of the 2021 nova eruption, which has parameters typical of the models of a very massive WD \citep[e.g.][]{2005ApJ...623..398Y}. However, the value of the mass accretion rate prior to the outburst derived in our models can be compared with the models of \citet{2005ApJ...623..398Y} with constant mass accretion rate before and after the outburst. From the fit we derived $\dot m$=3.3 $\times 10^{-9}$ M$_\odot$ yr$^{-1}$, but this does not justify the extremely rapid decrease by 3 visual magnitudes quoted above ($\simeq$ 1 day), which in \citet{2005ApJ...623..398Y}'s models turns out to be between 4.2 and 6 days (depending on the thermal state of teh WD) for $\dot m$ in the 10$^{-9}$-10$^{-8}$ M$_\odot$ yr$^{-1}$, even for the highest mass of their models' grid, 1.4 M$_\odot$. Although it is of course possible that the mass accretion rate was higher before the outburst, this results suggests that a different mechanism may have quickly depleted the envelope than the radiation pressure driven wind of \citet{2005ApJ...623..398Y}'s and other models.

The second model with which we obtained a spectral fit, Comptonization (Model B) does not allow to derive the WD mass and $\dot{M}$, so the jury is still out. This model would indicate instead that the accretion column may be sufficiently dense to be optically thick to Compton scattering ($\tau \approx 7$). In this scenario, the continuum shape is determined by the upscattering of seed photons rather than a simple distribution of plasma temperatures, implying that even 2.3 years post-outburst, the local mass accretion rate ($\dot{M}$) was high enough to maintain a high-density environment in the accretion funnel. According to this model, the lack of strong discrete emission lines in the spectrum is a consequence of the continuum's dominance and the scattering processes within the column. It is worth noting, however,  that the Comptonization-only scenario simplifies the physics by neglecting the direct competition with the bremsstrahlung processes. Classical studies \citep[e.g.,][]{Imamura1983ApJ...268..291I} have demonstrated that Compton scattering is significantly less dominant in these non-spherical geometries, primarily acting as a minor modifier to the primary thermal emission rather than dominating the continuum.   

While both models provide a statistically good fit to the quiescent spectra (\(\chi^2_{\nu} \approx 1.25\)), their physical implications differ significantly. The thermal model (Model A) indicates a high WD mass, which aligns with the fast optical decline of the nova, although with the data quality there is a degeneracy between temperature and metallicity. The Comptonization scenario (Model B) does not include all the physical phenomena, so it is a purely phenomenological description rather than a realistic physical alternative.

%
\subsection{The spin evolution} \label{sec:evol}
Previous works suggested that spin-up occurred in outburst (see Table \ref{tab:spin_evolution}), but in the {\it Chandra} light curves we find that spin-down may have started very shortly after the optical maximum. Even if the value $P_{spin} = 501.53 \pm 0.21$\,s has a large uncertainty, it  is larger than the two following spin estimates and overlaps them.
Instead, $P_{spin} = 501.31 \pm 0.19$\,s in the 2024 \xmm\ epoch implies a likely period decrease in quiescence. It is larger than the $501.72 \pm 0.11$\,s period measured with \chandra\ on day 37 of the outburst \citep{2021ApJ...922L..42D}. $\Delta P \approx -0.12$ s since day 37 implies post-outburst spin-up, possibly due to the accretion torques when the disk was re-established.
\begin{table*}
\centering
\caption{Evolution of the spin period ($P_{spin}$) of the white dwarf in \nova\ from 2018 to 2024.}
\label{tab:spin_evolution}
\begin{tabular}{lcccc}
\hline
\hline
Epoch & Source & Period ($P_{spin}$) [s] & $\Delta P$ from ZTF [s] & Physical State \\
\hline
2018--2021 & \cite{2021ATel14720....1M} & $501.4277 \pm 0.0004$ & --- & Deep Quiescence \\
2021 (Day 15) & \cite{2022ApJ...940L..56P} & $501.499 \pm 0.003$ & $+0.071$ & Post-Eruption \\
2021 (Day 28) & our re-analysis & $501.53 \pm 0.21$       &  $+0.1023$  & Post-Eruption  \\
2021 (Day 37) & \cite{Drake2022} & $501.72 \pm 0.11$ & $+0.292$ & Super-Soft Phase \\
2021 (Late) & \cite{Bhargava2024} & $501.52 \pm 0.04$ & $+0.092$ & Decline Phase \\
2022 & \cite{2022ApJ...940L..56P} & $501.40 \pm 0.02$ & $-0.028$ & Late Decline \\
2024 (June) & \citet{Luna2026V1405} & $501.328 \pm 0.024$ & $-0.099$ & Early Quiescence \\
2024 (Oct) & \textit{XMM-Newton/pn} (This work) & $501.31 \pm 0.19$ & $-0.117$ & Early Quiescence \\
\hline
\end{tabular}
\tablenotetext{a}{See Sections \ref{sec:lc} for a discussion about these periods.}
\end{table*}

Finally, the larger  modulation amplitude in the soft X-ray band, where the spectrum is dominated by the blackbody-like component, indicates that the origin in the heated accretion footprints on the white dwarf surface. In Fig. \ref{fig:lc}, we compare the 2024 \xmm\ light curve with the 2021 \chandra\ data, both phase-folded using the same recent ephemeris \citep[e.g.][]{2024CoSka..54b.128D}. 
Despite this phase migration and morphological evolution, the persistent detection of the spin signal confirms that the large-scale magnetic field of the white dwarf retained its structure throughout the eruption despite the immediate restart of the accretion flow as the system returned to quiescence.

\subsection{Localized burning ending the outburst}
By phase-folding both the 2021 \chandra\ and 2024 \xmm\ data with the same spin ephemeris, we observe a remarkable alignment of the pulse minima. This phase alignment was maintained despite the transition from nuclear-burning (SSS) to accretion-powered emission, with the latter being five orders of magnitudes less luminous. This clearly points out at the rotation, rather than stellar pulsations, as the root cause for the modulation in outburst.

The large luminosity of \nova\ during outburst, discussed in detail by \citet{Hachisu2026V1674her} proves unlikely a localized thermonuclear runaway as in \citet{Shara1982} and \citet{Orio1993}. However, the WD atmosphere was not homogeneous during the SSS phase \citep{Drake2022}. It is not unreasonable to think that this may have been due to localized burning under the polar caps at the end of the outburst, whether due to restarted accretion during which hydrogen was immediately burned under the polar caps, or to delayed turn-off in the polar regions. In the case of the intermediate polar novae V4743~Sgr and V407~Lup, \cite{Zemko2016} and \cite{Aydi2018} observed the emission measure of the SSS decreasing, while in turn the temperature was not immediately decreasing, apparently indicating a shrinking emission region rather than uniform cooling all over the WD surface.
\subsection{\nova\ as a fast magnetic rotator}
\citet{Drake2022} measured an increase in the period of the SSS modulation in the outburst and suggested that the expanding envelope had stopped corotating with the underlying WD during the initial outburst. However, when the SSS was observable,  the envelope may have been almost completely ejected. We suggest that the fast rotation coupled with a magnetic field above 10$^5$ Gauss is likely to have played the main role in mass ejection in this nova \citep[see][]{Orio1992}. \citet{Quimby2024} noticed that, although common envelope effects in short-period novae may cause an initial, a slow initial outflow as modeled by \citet{Shen2022} is not consistent with the sharp and abrupt rise of luminosity of \nova. In the MESA simulations of the outburst, the commonly assumed radiation pressure driven wind starts after an expansion to at least a few solar radii and when the effective temperature of the expanding envelope is below $\approx$150,000 K \citep{Shen2022}, but it starts at even with much lower  effective temperature and at a radius close to 100 R$_\odot$ in some previous models, depending on the system and WD parameters (D. Prialnik 2026, private communication). In any case, the expansion  of the envelope due to the TNR would have caused the radius to exceed  break-up velocity for \nova. However, \citet{Orio1992} found that, with a rotation velocity of $\geq$0.01 radians s$^{-1}$ (implying a rotation period of $\leq$628\,s) for any WD mass and magnetic field above 10$^5$ Gauss  - like an IP is likely to have - once even a small amount of mass loss starts occurring in the TNR, a {\it fast magnetic rotator wind} is launched inside the photosphere, depleting the accreted envelope quickly and efficiently.

The very rapid increase of luminosity observed by \citet{Quimby2024}, before the Roche Lobe could be filled, might be indeed explained by this mechanism. Even if only a little amount of mass is expelled explosively \citep{Starrfield2012, Starrfield2012b}, or if some mass is lost simply because the envelope cannot co-rotate with the WD core, the fast ``magnetic rotator wind'' was triggered. The rapid rise of \nova\ may be due to this different type of wind mechanism, as studied semi-analytically by \citet{Orio1992}. Thus we suggest that the fast rise and the rapid ``transition to an even faster rise'' observed by \citet{Quimby2024} was due to an initial explosive outflow, immediately followed by the fast magnetic rotator or centrifugal magnetic rotator wind. We may also hypothesize that two fast outflows with different velocities - an explosive ejection immediately followed by the fast magnetic rotator wind - may have caused the shock that accelerated particles, since gamma-rays were observed in coincidence with the optical peak \citep{2023MNRAS.521.5453S, Quimby2024, Aydi2025}.

In the model of \citet{Olbemo2026}, the magnetic field is unusually high, a few tens of MegaGauss. Their determination of course is model dependent and the parameters are - in our opinion - uncertain, but if this estimate is confirmed by other means, this nova was in the {\it centrifugal magnetical rotator} regime \cite{Orio1992}.
%
\section{Conclusions} \label{sec:conc}
We analyzed quasi-simultaneous \xmm\ and \nustar\ observations of the fast neon nova \nova\ obtained in October 2023, approximately 28 months post-outburst. Our main findings are as follows:

\begin{itemize}
    \item We found that the quiescent X-ray spectrum can be fitted either by a multi-temperature thermal plasma with $kT_{max} \geq 69$ keV or a Comptonized black-body component with an optical depth $\tau \approx 7$. Although the fit with the second model is statistically acceptable, we do not favor it because it does not include the fractional contribution of bremsstrahlung, which is an important source of cooling in the accretion column. The thermal plasma fit is consistent with a massive white dwarf ($M_{WD} \approx 1.35 M_\odot$) in an intermediate polar configuration. We also note that the conclusions by \citet{Olbemo2026} using the same data might be subject to specific modeling assumptions, some of which we discuss in Section 3.1.

    \item Timing analysis confirms a spin period of $501.35 \pm 0.19$\,s, with amplitude decreasing with increasing energy, as expected from energy-dependent absorption of post-shock emission in the accretion column. By re-analyzing early $Chandra$ data and including changes in the amplitude of the modulation calculating the period uncertainty, we found that a large period decrease on day 28 is not confirmed, although   two other measurements of spin-down in optical and X-rays during outburst indicate an actual period decrease by up to 0.3 s.

    \item By folding both the 2021 and 2024 light curves with the same spin ephemeris, we found that the pulse minima are aligned. This stability demonstrates that the magnetic poles were much hotter during the outburst, with a small migration with respect to the accretion footprints. This also suggests that the large-scale magnetosphere remained mostly intact throughout the eruption and return to quiescence. {\it Small differences in the period measurements indicates that the WD rotation was moderately slowed down after the outburst, and spun-up again by accretion after only a few months}.

    \item On the other hand, as in the cases of V4743~Sgr, V2491~Cyg, and V407~Lup, the persistent period of the modulation across different physical states (from the nuclear-burning SSS phase to accretion-powered quiescence) implies that rotation, rather than pulsations, is the origin of the observed periodicities even in the luminous SSS flux in the late phase of the outburst.

    \item We suggest that the WD in \nova\ {\it was either a fast magnetic rotator or a centrifugal magnetic rotator during the eruption}. Even a small amount of mass loss may have triggered this type of wind and the unusually rapid rise of the nova before a radiation pressure driven wind started, because of the rapid WD spin coupled with an IP magnetic field.
\end{itemize}


\section{Acknowledgments}

We acknowledge the anonymous referee for their careful reading and comments that significantly improved the manuscript. The lead author GJML is member of the CIC-CONICET (Argentina). 
The PI of the NuSTAR and XMM-Newton projects, M.O., has been funded by the NASA-Goddard NuSTAR grant 80NSSC23K1654 through the University of Wisconsin-Madison. AD and JM were funded by the EU ``NextGenerationEU Recovery and Resilience Plan'' for Slovakia under project No. 09I03-03-V04-00378. This research has made use of data obtained with the \nustar\ mission, a project led by the California Institute of Technology (Caltech), managed by the Jet Propulsion Laboratory (JPL) and funded by NASA. We used  the \nustar\ Data Analysis Software (NuSTARDAS) jointly developed by the ASI Science Data Center (ASDC, Italy) and the California Institute of Technology (Caltech, USA). XMM-Newton is an ESA science mission with instruments and contributions directly is funded by ESA Member States and NASA. We included observations done with the Samuel Oschin Telescope 48-inch and the 60-inch Telescope at the Palomar Observatory as part of the Zwicky Transient Facility project. ZTF is supported by the National Science Foundation under Grants No. AST-1440341 and AST-2034437 and a collaboration including current partners Caltech, IPAC, the Oskar Klein Center at Stockholm University, the University of Maryland, University of California, Berkeley, the University of Wisconsin at Milwaukee, University of Warwick, Ruhr University, Cornell University, Northwestern University and Drexel University. Operations are conducted by COO, IPAC, and UW.

%

\vspace{5mm}
\facilities{NuSTAR, \xmm, $ZTF$}






\bibliography{listaref_MASTER}{}

@ARTICLE{2005ApJ...623..398Y,
   author = {{Yaron}, O. and {Prialnik}, D. and {Shara}, M.~M. and {Kovetz}, A.
	},
    title = "{An Extended Grid of Nova Models. II. The Parameter Space of Nova Outbursts}",
  journal = {\apj},
   eprint = {astro-ph/0503143},
     year = 2005,
    month = apr,
   volume = 623,
    pages = {398-410},
      doi = {10.1086/428435},
   adsurl = {http://adsabs.harvard.edu/abs/2005ApJ...623..398Y}
}

@ARTICLE{2000ApJ...542..914W,
   author = {{Wilms}, J. and {Allen}, A. and {McCray}, R.},
    title = "{On the Absorption of X-Rays in the Interstellar Medium}",
  journal = {\apj},
   eprint = {astro-ph/0008425},
     year = 2000,
    month = oct,
   volume = 542,
    pages = {914-924},
      doi = {10.1086/317016},
   adsurl = {http://adsabs.harvard.edu/abs/2000ApJ...542..914W}
}

@ARTICLE{2017PASP..129f2001M,
   author = {{Mukai}, K.},
    title = "{X-Ray Emissions from Accreting White Dwarfs: A Review}",
  journal = {\pasp},
archivePrefix = "arXiv",
   eprint = {1703.06171},
 primaryClass = "astro-ph.HE",
     year = 2017,
    month = jun,
   volume = 129,
   number = 6,
    pages = {062001},
      doi = {10.1088/1538-3873/aa6736},
   adsurl = {http://adsabs.harvard.edu/abs/2017PASP..129f2001M}
}

@ARTICLE{2022ApJ...940L..56P,
       author = {{Patterson}, Joseph and {Enenstein}, Josie and {de Miguel}, Enrique and {Epstein-Martin}, Marguerite and {Kemp}, Jonathan and {Sabo}, Richard and {Cooney}, Walt and {Vanmunster}, Tonny and {Dubovsky}, Pavol and {Hambsch}, Franz-Josef and {Myers}, Gordon and {Lemay}, Damien and {Sokolovsky}, Kirill and {Collins}, Donald and {Campbell}, Tut and {Roberts}, George and {Richmond}, Michael and {Brincat}, Stephen and {Ulowetz}, Joseph and {Dvorak}, Shawn and {Tordai}, Tam{\'a}s and {Dufoer}, Sjoerd and {Cahaly}, Andrew and {Galdies}, Charles and {Goff}, Bill and {Wilkin}, Francis P. and {Wood}, Matt A.},
        title = "{The Periodic Signals of Nova V1674 Herculis (2021)}",
      journal = {\apjl},
         year = 2022,
        month = dec,
       volume = {940},
       number = {2},
          eid = {L56},
        pages = {L56},
          doi = {10.3847/2041-8213/ac9ebe},
archivePrefix = {arXiv},
       eprint = {2207.00181},
 primaryClass = {astro-ph.SR},
       adsurl = {https://ui.adsabs.harvard.edu/abs/2022ApJ...940L..56P}
}

@ARTICLE{2021ApJ...922L..42D,
       author = {{Drake}, Jeremy J. and {Ness}, Jan-Uwe and {Page}, Kim L. and {Luna}, G.~J.~M. and {Beardmore}, Andrew P. and {Orio}, Marina and {Osborne}, Julian P. and {Mr{\'o}z}, Przemek and {Starrfield}, Sumner and {Banerjee}, Dipankar P.~K. and {Balman}, Solen and {Darnley}, M.~J. and {Bhargava}, Y. and {Dewangan}, G.~C. and {Singh}, K.~P.},
        title = "{The Remarkable Spin-down and Ultrafast Outflows of the Highly Pulsed Supersoft Source of Nova Herculis 2021}",
      journal = {\apjl},
         year = 2021,
        month = dec,
       volume = {922},
       number = {2},
          eid = {L42},
        pages = {L42},
          doi = {10.3847/2041-8213/ac34fd},
archivePrefix = {arXiv},
       eprint = {2110.14058},
 primaryClass = {astro-ph.HE},
       adsurl = {https://ui.adsabs.harvard.edu/abs/2021ApJ...922L..42D}
}

@ARTICLE{2021ATel14720....1M,
       author = {{Mroz}, P. and {Burdge}, K. and {Roestel}, J. van and {Prince}, T. and {Kong}, A.~K.~H. and {Li}, K. -L.},
        title = "{An 8.4 min period in the archival ZTF light curve of Nova Herculis 2021}",
      journal = {ATel},
         year = 2021,
        month = jun,
       volume = {14720},
        pages = {1},
       adsurl = {https://ui.adsabs.harvard.edu/abs/2021ATel14720....1M}
}

@ARTICLE{2021ATel14835....1S,
       author = {{Shugarov}, Sergey and {Afonina}, Marina},
        title = "{Detection of a possibly orbital period of the classical Nova V1674 Her}",
      journal = {ATel},
         year = 2021,
        month = aug,
       volume = {14835},
        pages = {1},
       adsurl = {https://ui.adsabs.harvard.edu/abs/2021ATel14835....1S}
}

@ARTICLE{2021ATel14856....1P,
       author = {{Patterson}, J. and {Epstein-Martin}, M. and {Vanmunster}, T. and {Kemp}, J.},
        title = "{The Orbital and Pulse Periods of V1674 Herculis}",
      journal = {ATel},
         year = 2021,
        month = aug,
       volume = {14856},
        pages = {1},
       adsurl = {https://ui.adsabs.harvard.edu/abs/2021ATel14856....1P}
}

@ARTICLE{2023MNRAS.521.5453S,
       author = {{Sokolovsky}, K.~V. and {Johnson}, T.~J. and {Buson}, S. and {Jean}, P. and {Cheung}, C.~C. and {Mukai}, K. and {Chomiuk}, L. and {Aydi}, E. and {Molina}, B. and {Kawash}, A. and {Linford}, J.~D. and {Mioduszewski}, A.~J. and {Rupen}, M.~P. and {Sokoloski}, J.~L. and {Williams}, M.~N. and {Steinberg}, E. and {Vurm}, I. and {Metzger}, B.~D. and {Page}, K.~L. and {Orio}, M. and {Quimby}, R.~M. and {Shafter}, A.~W. and {Corbett}, H. and {Bolzoni}, S. and {DeYoung}, J. and {Menzies}, K. and {Romanov}, F.~D. and {Richmond}, M. and {Ulowetz}, J. and {Vanmunster}, T. and {Williamson}, G. and {Lane}, D.~J. and {Bartnik}, M. and {Bellaver}, M. and {Bruinsma}, E. and {Dugan}, E. and {Fedewa}, J. and {Gerhard}, C. and {Painter}, S. and {Peterson}, D. -M. and {Rodriguez}, J.~E. and {Smith}, C. and {Sullivan}, H. and {Watson}, S.},
        title = "{The multiwavelength view of shocks in the fastest nova V1674 Her}",
      journal = {\mnras},
         year = 2023,
        month = jun,
       volume = {521},
       number = {4},
        pages = {5453-5472},
          doi = {10.1093/mnras/stad887},
archivePrefix = {arXiv},
       eprint = {2302.03043},
 primaryClass = {astro-ph.HE},
       adsurl = {https://ui.adsabs.harvard.edu/abs/2023MNRAS.521.5453S}
}

@ARTICLE{Orio2022,
       author = {{Orio}, M. and {Gendreau}, K. and {Giese}, M. and {Luna}, G.~J.~M. and {Magdolen}, J. and {Pei}, S. and {Sun}, B. and {Behar}, E. and {Dobrotka}, A. and {Mikolajewska}, J. and {Pasham}, Dheeraj R. and {Strohmayer}, T.~E.},
        title = "{NICER Monitoring of Supersoft X-Ray Sources}",
      journal = {\apj},
         year = 2022,
        month = jun,
       volume = {932},
       number = {1},
          eid = {45},
        pages = {45},
          doi = {10.3847/1538-4357/ac63be},
archivePrefix = {arXiv},
       eprint = {2204.01660},
 primaryClass = {astro-ph.HE},
       adsurl = {https://ui.adsabs.harvard.edu/abs/2022ApJ...932...45O}
}

@ARTICLE{2021ApJ...911...80M,
       author = {{Maiolino}, T. and {Titarchuk}, L. and {Wang}, W. and {Frontera}, F. and {Orlandini}, M.},
        title = "{Comptonization as an Origin of the Continuum in Intermediate Polars}",
      journal = {\apj},
         year = 2021,
        month = apr,
       volume = {911},
       number = {2},
          eid = {80},
        pages = {80},
          doi = {10.3847/1538-4357/abe9bb},
archivePrefix = {arXiv},
       eprint = {2102.11789},
 primaryClass = {astro-ph.HE},
       adsurl = {https://ui.adsabs.harvard.edu/abs/2021ApJ...911...80M}
}

@ARTICLE{Quimby2024,
       author = {{Quimby}, Robert M. and {Metzger}, Brian D. and {Shen}, Ken J. and {Shafter}, Allen W. and {Corbett}, Hank and {Overton}, Madeline},
        title = "{The Rise of Nova V1674 Herculis}",
      journal = {\apj},
         year = 2024,
        month = dec,
       volume = {977},
       number = {1},
          eid = {17},
        pages = {17},
          doi = {10.3847/1538-4357/ad887f},
archivePrefix = {arXiv},
       eprint = {2410.16460},
 primaryClass = {astro-ph.HE},
       adsurl = {https://ui.adsabs.harvard.edu/abs/2024ApJ...977...17Q}
}

@ARTICLE{2024MNRAS.527.1405H,
       author = {{Habtie}, Gesesew R. and {Das}, Ramkrishna and {Pandey}, Ruchi and {Ashok}, N.~M. and {Dubovsky}, Pavol A.},
        title = "{Study of the fastest classical nova, V1674 Her: photoionization and morpho-kinemetic model analysis}",
      journal = {\mnras},
         year = 2024,
        month = jan,
       volume = {527},
       number = {1},
        pages = {1405-1423},
          doi = {10.1093/mnras/stad3295},
archivePrefix = {arXiv},
       eprint = {2310.16619},
 primaryClass = {astro-ph.SR},
       adsurl = {https://ui.adsabs.harvard.edu/abs/2024MNRAS.527.1405H}
}

@ARTICLE{2024CoSka..54b.128D,
       author = {{Dubovsk{\'y}}, P.~A. and {Petr{\'\i}k}, K. and {Breus}, V.},
        title = "{Nova Herculi 2021 as an intermediate polar}",
      journal = {Contributions of the Astronomical Observatory Skalnate Pleso},
         year = 2024,
        month = feb,
       volume = {54},
       number = {2},
        pages = {128-134},
          doi = {10.31577/caosp.2024.54.2.128},
       adsurl = {https://ui.adsabs.harvard.edu/abs/2024CoSka..54b.128D}
}

@ARTICLE{Worley2025,
       author = {{Worley}, John T. and {Orio}, Marina and {Dobrotka}, Andrej and {Magdolen}, Jozef and {Page}, Kim and {Behar}, Ehud and {Drake}, Jeremy J. and {Mitrani}, Sharon},
        title = "{X-Ray Observations of Nova Scorpii 2023 (V1716 Sco) in Outburst}",
      journal = {\apj},
         year = 2025,
        month = dec,
       volume = {995},
       number = {1},
          eid = {30},
        pages = {30},
          doi = {10.3847/1538-4357/ae157b},
archivePrefix = {arXiv},
       eprint = {2510.19102},
 primaryClass = {astro-ph.HE},
       adsurl = {https://ui.adsabs.harvard.edu/abs/2025ApJ...995...30W}
}

@ARTICLE{Zemko2016,
       author = {{Zemko}, P. and {Orio}, M. and {Mukai}, K. and {Bianchini}, A. and {Ciroi}, S. and {Cracco}, V.},
        title = "{V4743 Sgr, a magnetic nova?}",
      journal = {\mnras},
         year = 2016,
        month = aug,
       volume = {460},
       number = {3},
        pages = {2744-2751},
          doi = {10.1093/mnras/stw1199},
archivePrefix = {arXiv},
       eprint = {1606.00225},
 primaryClass = {astro-ph.SR},
       adsurl = {https://ui.adsabs.harvard.edu/abs/2016MNRAS.460.2744Z}
}

@article{Dobrotka2017,
  author         = {Dobrotka, A. and Ness, J. -U. and Zemko, P. and Orio, M.},
  title          = {{The long-term X-ray behaviour of the old nova V4743 Sagittarii}},
  journal        = {\mnras},
  year           = {2017},
  month          = {nov},
  volume         = {471},
  number         = {4},
  pages          = {4041-4048},
  doi            = {10.1093/mnras/stx1887},
  archiveprefix  = {arXiv},
  eprint         = {1707.07994},
  primaryclass   = {astro-ph.SR}
}

@article{Zemko2015,
  author         = {Zemko, P. and Orio, M. and Shugarov, S.},
  title          = {{X-ray analysis of the recurrent nova V2491 Cyg}},
  journal        = {\mnras},
  year           = {2015},
  month          = {jan},
  volume         = {446},
  number         = {2},
  pages          = {1343-1349},
  doi            = {10.1093/mnras/stu2208},
  archiveprefix  = {arXiv},
  eprint         = {1410.6380},
  primaryclass   = {astro-ph.SR}
}

@ARTICLE{Aydi2018,
       author = {{Aydi}, E. and {Orio}, M. and {Beardmore}, A.~P. and {Ness}, J.-U. and {Page}, K.~L. and {Kuin}, N.~P.~M. and {Walter}, F.~M. and {Buckley}, D.~A.~H. and {Mohamed}, S. and {Whitelock}, P. and {Osborne}, J.~P. and {Strader}, J. and {Chomiuk}, L. and {Darnley}, M.~J. and {Dobrotka}, A. and {Kniazev}, A. and {Miszalski}, B. and {Myers}, G. and {Ospina}, N. and {Henze}, M. and {Starrfield}, S. and {Woodward}, C.~E.},
        title = "{Multiwavelength observations of V407 Lupi (ASASSN-16kt) - a very fast nova erupting in an intermediate polar}",
      journal = {\mnras},
         year = 2018,
        month = oct,
       volume = {480},
       number = {1},
        pages = {572-609},
          doi = {10.1093/mnras/sty1759},
archivePrefix = {arXiv},
       eprint = {1807.00706},
 primaryClass = {astro-ph.SR},
       adsurl = {https://ui.adsabs.harvard.edu/abs/2018MNRAS.480..572A}
}

@ARTICLE{Orio2024,
       author = {{Orio}, M. and {Melicher{\v{c}}{\'\i}k}, M. and {Ciroi}, S. and {Canton}, V. and {Aydi}, E. and {Buckley}, D.~A.~H. and {Dobrotka}, A. and {Luna}, G.~J.~M. and {Ness}, J.},
        title = "{V407 Lup, an intermediate polar nova}",
      journal = {\mnras},
         year = 2024,
        month = sep,
       volume = {533},
       number = {2},
        pages = {1541-1549},
          doi = {10.1093/mnras/stae1922},
archivePrefix = {arXiv},
       eprint = {2408.03779},
 primaryClass = {astro-ph.HE},
       adsurl = {https://ui.adsabs.harvard.edu/abs/2024MNRAS.533.1541O}
}

@article{Drake2022,
  author         = {Drake, J. J. and Ness, J. -U. and Page, K. L. and Luna, G. J. M. and Beardmore, A. P. and Orio, M. and Starrfield, S.},
  title          = {{X-Ray Grating Spectroscopy of the Fast Nova V1674 Herculis in the Super Soft Source Phase}},
  journal        = {\apjl},
  year           = {2022},
  month          = {jan},
  volume         = {925},
  number         = {2},
  eid            = {L23},
  pages          = {L23},
  doi            = {10.3847/2041-8213/ac4b07},
  archiveprefix  = {arXiv},
  eprint         = {2112.03932},
  primaryclass   = {astro-ph.HE}
}

@ARTICLE{Bhargava2024,
       author = {{Bhargava}, Yash and {Dewangan}, Gulab Chand and {Anupama}, G.~C. and {Kamath}, U.~S. and {Sonith}, L.~S. and {Singh}, Kulinder Pal and {Drake}, J.~J. and {Beardmore}, A. and {Luna}, G.~J.~M. and {Orio}, M. and {Page}, K.~L.},
        title = "{Soft X-ray and FUV observations of Nova Her 2021 (V1674 Her) with AstroSat}",
      journal = {\mnras},
         year = 2024,
        month = feb,
       volume = {528},
       number = {1},
        pages = {28-38},
          doi = {10.1093/mnras/stad3870},
archivePrefix = {arXiv},
       eprint = {2312.10291},
 primaryClass = {astro-ph.HE},
       adsurl = {https://ui.adsabs.harvard.edu/abs/2024MNRAS.528...28B}
}

@ARTICLE{Astropy2022,
       author = {{Astropy Collaboration} and {Price-Whelan}, Adrian M. and {Lim}, Pey Lian and {Earl}, Nicholas and {Starkman}, Nathaniel and {Bradley}, Larry and {Shupe}, David L. and {Patil}, Aarya A. and {Corrales}, Lia and {Brasseur}, C.~E. and {N{\"o}the}, Maximilian and {Donath}, Axel and {Tollerud}, Erik and {Morris}, Brett M. and {Ginsburg}, Adam and {Vaher}, Eero and {Weaver}, Benjamin A. and {Tocknell}, James and {Jamieson}, William and {van Kerkwijk}, Marten H. and {Robitaille}, Thomas P. and {Merry}, Bruce and {Bachetti}, Matteo and {G{\"u}nther}, H. Moritz and {Aldcroft}, Thomas L. and {Alvarado-Montes}, Jaime A. and {Archibald}, Anne M. and {B{\'o}di}, Attila and {Bapat}, Shreyas and {Barentsen}, Geert and {Baz{\'a}n}, Juanjo and {Biswas}, Manish and {Boquien}, M{\'e}d{\'e}ric and {Burke}, D.~J. and {Cara}, Daria and {Cara}, Mihai and {Conroy}, Kyle E. and {Conseil}, Simon and {Craig}, Matthew W. and {Cross}, Robert M. and {Cruz}, Kelle L. and {D'Eugenio}, Francesco and {Dencheva}, Nadia and {Devillepoix}, Hadrien A.~R. and {Dietrich}, J{\"o}rg P. and {Eigenbrot}, Arthur Davis and {Erben}, Thomas and {Ferreira}, Leonardo and {Foreman-Mackey}, Daniel and {Fox}, Ryan and {Freij}, Nabil and {Garg}, Suyog and {Geda}, Robel and {Glattly}, Lauren and {Gondhalekar}, Yash and {Gordon}, Karl D. and {Grant}, David and {Greenfield}, Perry and {Groener}, Austen M. and {Guest}, Steve and {Gurovich}, Sebastian and {Handberg}, Rasmus and {Hart}, Akeem and {Hatfield-Dodds}, Zac and {Homeier}, Derek and {Hosseinzadeh}, Griffin and {Jenness}, Tim and {Jones}, Craig K. and {Joseph}, Prajwel and {Kalmbach}, J. Bryce and {Karamehmetoglu}, Emir and {Ka{\l}uszy{\'n}ski}, Miko{\l}aj and {Kelley}, Michael S.~P. and {Kern}, Nicholas and {Kerzendorf}, Wolfgang E. and {Koch}, Eric W. and {Kulumani}, Shankar and {Lee}, Antony and {Ly}, Chun and {Ma}, Zhiyuan and {MacBride}, Conor and {Maljaars}, Jakob M. and {Muna}, Demitri and {Murphy}, N.~A. and {Norman}, Henrik and {O'Steen}, Richard and {Oman}, Kyle A. and {Pacifici}, Camilla and {Pascual}, Sergio and {Pascual-Granado}, J. and {Patil}, Rohit R. and {Perren}, Gabriel I. and {Pickering}, Timothy E. and {Rastogi}, Tanuj and {Roulston}, Benjamin R. and {Ryan}, Daniel F. and {Rykoff}, Eli S. and {Sabater}, Jose and {Sakurikar}, Parikshit and {Salgado}, Jes{\'u}s and {Sanghi}, Aniket and {Saunders}, Nicholas and {Savchenko}, Volodymyr and {Schwardt}, Ludwig and {Seifert-Eckert}, Michael and {Shih}, Albert Y. and {Jain}, Anany Shrey and {Shukla}, Gyanendra and {Sick}, Jonathan and {Simpson}, Chris and {Singanamalla}, Sudheesh and {Singer}, Leo P. and {Singhal}, Jaladh and {Sinha}, Manodeep and {Sip{\H{o}}cz}, Brigitta M. and {Spitler}, Lee R. and {Stansby}, David and {Streicher}, Ole and {{\v{S}}umak}, Jani and {Swinbank}, John D. and {Taranu}, Dan S. and {Tewary}, Nikita and {Tremblay}, Grant R. and {de Val-Borro}, Miguel and {Van Kooten}, Samuel J. and {Vasovi{\'c}}, Zlatan and {Verma}, Shresth and {de Miranda Cardoso}, Jos{\'e} Vin{\'\i}cius and {Williams}, Peter K.~G. and {Wilson}, Tom J. and {Winkel}, Benjamin and {Wood-Vasey}, W.~M. and {Xue}, Rui and {Yoachim}, Peter and {Zhang}, Chen and {Zonca}, Andrea and {Astropy Project Contributors}},
        title = "{The Astropy Project: Sustaining and Growing a Community-oriented Open-source Project and the Latest Major Release (v5.0) of the Core Package}",
      journal = {\apj},
         year = 2022,
        month = aug,
       volume = {935},
       number = {2},
          eid = {167},
        pages = {167},
          doi = {10.3847/1538-4357/ac7c74},
archivePrefix = {arXiv},
       eprint = {2206.14220},
 primaryClass = {astro-ph.IM},
       adsurl = {https://ui.adsabs.harvard.edu/abs/2022ApJ...935..167A}
}

@ARTICLE{Luna2025tess,
       author = {{Luna}, G.~J.~M. and {Dobrotka}, A. and {Orio}, M.},
        title = "{A TESS View of Post-Eruption Variability in the Novae V1405 Cas,V1716 Sco, and V1674 Her}",
      journal = {arXiv e-prints},
         year = 2025,
        month = nov,
          eid = {arXiv:2511.06399},
        pages = {arXiv:2511.06399},
          doi = {10.48550/arXiv.2511.06399},
archivePrefix = {arXiv},
       eprint = {2511.06399},
 primaryClass = {astro-ph.SR},
       adsurl = {https://ui.adsabs.harvard.edu/abs/2025arXiv251106399L}
}

@ARTICLE{Hachisu2026V1674her,
       author = {{Hachisu}, Izumi and {Kato}, Mariko},
        title = "{Formation of an Optically Thick Shocked Shell in the Very Fast Nova V1674 Herculis: The Origin of Superbrightness}",
      journal = {\apj},
         year = 2026,
        month = feb,
       volume = {998},
       number = {1},
          eid = {156},
        pages = {156},
          doi = {10.3847/1538-4357/ae314b},
archivePrefix = {arXiv},
       eprint = {2512.20175},
 primaryClass = {astro-ph.SR},
       adsurl = {https://ui.adsabs.harvard.edu/abs/2026ApJ...998..156H}
}

@ARTICLE{Shara1982,
       author = {{Shara}, M.~M.},
        title = "{Localized thermonuclear runaways and volcanoes on degenerate dwarf stars}",
      journal = {\apj},
         year = 1982,
        month = oct,
       volume = {261},
        pages = {649-660},
          doi = {10.1086/160376},
       adsurl = {https://ui.adsabs.harvard.edu/abs/1982ApJ...261..649S}
}

@ARTICLE{Aydi2025,
       author = {{Aydi}, Elias and {Monnier}, John D. and {M{\'e}rand}, Antoine and {Schaefer}, Gail H. and {Chomiuk}, Laura and {Otulakowska-Hypka}, Magdalena and {Fan}, Jhih-Ling and {Li}, Kwan Lok and {Sokolovsky}, Kirill V. and {Salinas}, Ricardo and {Tucker}, Michael and {Shappee}, Benjamin and {Rudy}, Richard and {Page}, Kim L. and {Kuin}, N. Paul M. and {Buckley}, David A.~H. and {Craig}, Peter and {Izzo}, Luca and {Linford}, Justin and {Metzger}, Brian D. and {Mukai}, Koji and {Orio}, Marina and {Shen}, Ken J. and {Strader}, Jay and {Sokoloski}, Jennifer L. and {Williams}, Robert E. and {Williams}, Montana N. and {Habtie}, Gesesew R. and {Kraus}, Stefan and {Anugu}, Narsireddy and {Bouquin}, Jean-Baptiste Le and {Chhabra}, Sorabh and {Codron}, Isabelle and {Gardner}, Tyler and {Gutierrez}, Mayra and {Ibrahim}, Noura and {Lanthermann}, Cyprien and {Setterholm}, Benjamin R. and {Ashall}, Christopher and {Hinkle}, Jason T. and {Jaeger}, Thomas de and {Payne}, Anna V.},
        title = "{Multiple outflows and delayed ejections revealed by early imaging of novae}",
      journal = {Nature Astronomy},
         year = 2025,
        month = dec,
          doi = {10.1038/s41550-025-02725-1},
archivePrefix = {arXiv},
       eprint = {2512.05220},
 primaryClass = {astro-ph.HE},
       adsurl = {https://ui.adsabs.harvard.edu/abs/2025NatAs.tmp..249A}
}

@ARTICLE{Wagner2021,
       author = {{Wagner}, R.~M. and {Woodward}, C.~E. and {Starrfield}, S. and {Banerjee}, D.~P.~K. and {Evans}, A.},
        title = "{V1674 Her is a Neon Nova}",
      journal = {ATel},
         year = 2021,
        month = jul,
       volume = {14746},
        pages = {1},
       adsurl = {https://ui.adsabs.harvard.edu/abs/2021ATel14746....1W}
}

@ARTICLE{Maccarone2021,
       author = {{Maccarone}, Thomas J. and {Beardmore}, Andy and {Mukai}, Koji and {Page}, Kim and {Pichardo Marcano}, Manuel and {Rivera Sandoval}, Liliana},
        title = "{X-ray Pulsations from Nova Her 2021}",
      journal = {The Astronomer's Telegram},
         year = 2021,
        month = jul,
       volume = {14776},
        pages = {1},
       adsurl = {https://ui.adsabs.harvard.edu/abs/2021ATel14776....1M}
}

@ARTICLE{Orio1992,
       author = {{Orio}, M. and {Trussoni}, E. and {Oegelman}, H.},
        title = "{Classical novae as fast magnetic rotators.}",
      journal = {\aap},
         year = 1992,
        month = apr,
       volume = {257},
        pages = {548-556},
       adsurl = {https://ui.adsabs.harvard.edu/abs/1992A&A...257..548O}
}

@ARTICLE{Orio1993,
       author = {{Orio}, Marina and {Shaviv}, Giora},
        title = "{Local Thermonuclear Runaways among Classical Novae}",
      journal = {\apss},
         year = 1993,
        month = apr,
       volume = {202},
       number = {2},
        pages = {273-288},
          doi = {10.1007/BF00626882},
       adsurl = {https://ui.adsabs.harvard.edu/abs/1993Ap&SS.202..273O}
}

@ARTICLE{Shen2022,
       author = {{Shen}, Ken J. and {Quataert}, Eliot},
        title = "{Binary Interaction Dominates Mass Ejection in Classical Novae}",
      journal = {\apj},
         year = 2022,
        month = oct,
       volume = {938},
       number = {1},
          eid = {31},
        pages = {31},
          doi = {10.3847/1538-4357/ac9136},
archivePrefix = {arXiv},
       eprint = {2205.06283},
 primaryClass = {astro-ph.SR},
       adsurl = {https://ui.adsabs.harvard.edu/abs/2022ApJ...938...31S}
}

@ARTICLE{Starrfield2012b,  
       author = {{Starrfield}, S. and {Timmes}, F.~X. and {Iliadis}, C. and {Hix}, W.~R. and {Arnett}, W.~D. and {Meakin}, C. and {Sparks}, W.~M.},
        title = "{Hydrodynamic Studies of the Evolution of Recurrent, Symbiotic and Dwarf Novae: the White Dwarf Components are Growing in Mass}",
      journal = {Baltic Astronomy},
         year = 2012,
        month = jan,
       volume = {21},
        pages = {76-87},
          doi = {10.1515/astro-2017-0361},
archivePrefix = {arXiv},
       eprint = {1211.6145},
 primaryClass = {astro-ph.SR},
       adsurl = {https://ui.adsabs.harvard.edu/abs/2012BaltA..21...76S}
}

@ARTICLE{Starrfield2012,        
       author = {{Starrfield}, S. and {Iliadis}, C. and {Timmes}, F.~X. and {Hix}, W.~R. and {Arnett}, W.~D. and {Meakin}, C. and {Sparks}, W.~M.},
        title = "{Theoretical studies of accretion of matter onto white dwarfs and the single degenerate scenario for supernovae of Type Ia}",
      journal = {Bulletin of the Astronomical Society of India},
         year = 2012,
        month = sep,
       volume = {40},
        pages = {419},
          doi = {10.48550/arXiv.1210.6086},
archivePrefix = {arXiv},
       eprint = {1210.6086},
 primaryClass = {astro-ph.SR},
       adsurl = {https://ui.adsabs.harvard.edu/abs/2012BASI...40..419S}
}

@ARTICLE{Graham2019,
       author = {{Graham}, Matthew J. and {Kulkarni}, S.~R. and {Bellm}, Eric C. and {Adams}, Scott M. and {Barbarino}, Cristina and {Blagorodnova}, Nadejda and {Bodewits}, Dennis and {Bolin}, Bryce and {Brady}, Patrick R. and {Cenko}, S. Bradley and {Chang}, Chan-Kao and {Coughlin}, Michael W. and {De}, Kishalay and {Eadie}, Gwendolyn and {Farnham}, Tony L. and {Feindt}, Ulrich and {Franckowiak}, Anna and {Fremling}, Christoffer and {Gezari}, Suvi and {Ghosh}, Shaon and {Goldstein}, Daniel A. and {Golkhou}, V. Zach and {Goobar}, Ariel and {Ho}, Anna Y.~Q. and {Huppenkothen}, Daniela and {Ivezi{\'c}}, {\v{Z}}eljko and {Jones}, R. Lynne and {Juric}, Mario and {Kaplan}, David L. and {Kasliwal}, Mansi M. and {Kelley}, Michael S.~P. and {Kupfer}, Thomas and {Lee}, Chien-De and {Lin}, Hsing Wen and {Lunnan}, Ragnhild and {Mahabal}, Ashish A. and {Miller}, Adam A. and {Ngeow}, Chow-Choong and {Nugent}, Peter and {Ofek}, Eran O. and {Prince}, Thomas A. and {Rauch}, Ludwig and {van Roestel}, Jan and {Schulze}, Steve and {Singer}, Leo P. and {Sollerman}, Jesper and {Taddia}, Francesco and {Yan}, Lin and {Ye}, Quan-Zhi and {Yu}, Po-Chieh and {Barlow}, Tom and {Bauer}, James and {Beck}, Ron and {Belicki}, Justin and {Biswas}, Rahul and {Brinnel}, Valery and {Brooke}, Tim and {Bue}, Brian and {Bulla}, Mattia and {Burruss}, Rick and {Connolly}, Andrew and {Cromer}, John and {Cunningham}, Virginia and {Dekany}, Richard and {Delacroix}, Alex and {Desai}, Vandana and {Duev}, Dmitry A. and {Feeney}, Michael and {Flynn}, David and {Frederick}, Sara and {Gal-Yam}, Avishay and {Giomi}, Matteo and {Groom}, Steven and {Hacopians}, Eugean and {Hale}, David and {Helou}, George and {Henning}, John and {Hover}, David and {Hillenbrand}, Lynne A. and {Howell}, Justin and {Hung}, Tiara and {Imel}, David and {Ip}, Wing-Huen and {Jackson}, Edward and {Kaspi}, Shai and {Kaye}, Stephen and {Kowalski}, Marek and {Kramer}, Emily and {Kuhn}, Michael and {Landry}, Walter and {Laher}, Russ R. and {Mao}, Peter and {Masci}, Frank J. and {Monkewitz}, Serge and {Murphy}, Patrick and {Nordin}, Jakob and {Patterson}, Maria T. and {Penprase}, Bryan and {Porter}, Michael and {Rebbapragada}, Umaa and {Reiley}, Dan and {Riddle}, Reed and {Rigault}, Mickael and {Rodriguez}, Hector and {Rusholme}, Ben and {van Santen}, Jakob and {Shupe}, David L. and {Smith}, Roger M. and {Soumagnac}, Maayane T. and {Stein}, Robert and {Surace}, Jason and {Szkody}, Paula and {Terek}, Scott and {Van Sistine}, Angela and {van Velzen}, Sjoert and {Vestrand}, W. Thomas and {Walters}, Richard and {Ward}, Charlotte and {Zhang}, Chaoran and {Zolkower}, Jeffry},
        title = "{The Zwicky Transient Facility: Science Objectives}",
      journal = {\pasp},
         year = 2019,
        month = jul,
       volume = {131},
       number = {1001},
        pages = {078001},
          doi = {10.1088/1538-3873/ab006c},
archivePrefix = {arXiv},
       eprint = {1902.01945},
 primaryClass = {astro-ph.IM},
       adsurl = {https://ui.adsabs.harvard.edu/abs/2019PASP..131g8001G}
}

@ARTICLE{Olbemo2026,
       author = {{Olbemo}, Tekeba and {Errando}, Manel and {Gokus}, Andrea},
        title = "{X-ray spectroscopy mass constraints on V1674 Her: the fastest nova does not have a near-Chandrasekhar white dwarf}",
      journal = {arXiv e-prints},
         year = 2026,
        month = may,
          eid = {arXiv:2605.06917},
        pages = {arXiv:2605.06917},
archivePrefix = {arXiv},
       eprint = {2605.06917},
 primaryClass = {astro-ph.HE},
       adsurl = {https://ui.adsabs.harvard.edu/abs/2026arXiv260506917O}
}

@ARTICLE{Suleimanov2025,
       author = {{Suleimanov}, V.~F. and {Ducci}, L. and {Doroshenko}, V. and {Werner}, K.},
        title = "{X-ray spectroscopy method of white dwarf mass determination in intermediate polars: External systematic uncertainties}",
      journal = {\aap},
         year = 2025,
        month = aug,
       volume = {700},
          eid = {A180},
        pages = {A180},
          doi = {10.1051/0004-6361/202554368},
archivePrefix = {arXiv},
       eprint = {2506.03711},
 primaryClass = {astro-ph.SR},
       adsurl = {https://ui.adsabs.harvard.edu/abs/2025A&A...700A.180S}
}

@ARTICLE{Luna2015A&A...578A..15L,
       author = {{Luna}, G.~J.~M. and {Raymond}, J.~C. and {Brickhouse}, N.~S. and {Mauche}, C.~W. and {Suleimanov}, V.},
        title = "{Testing the cooling flow model in the intermediate polar EX Hydrae}",
      journal = {\aap},
         year = 2015,
        month = jun,
       volume = {578},
          eid = {A15},
        pages = {A15},
          doi = {10.1051/0004-6361/201525755},
archivePrefix = {arXiv},
       eprint = {1504.01342},
 primaryClass = {astro-ph.HE},
       adsurl = {https://ui.adsabs.harvard.edu/abs/2015A&A...578A..15L}
}

@ARTICLE{stellingwerf1978,
       author = {{Stellingwerf}, R.~F.},
        title = "{Period determination using phase dispersion minimization.}",
      journal = {\apj},
         year = 1978,
        month = sep,
       volume = {224},
        pages = {953-960},
          doi = {10.1086/156444},
       adsurl = {https://ui.adsabs.harvard.edu/abs/1978ApJ...224..953S}
}

@INPROCEEDINGS{mkcflow1988,
       author = {{Mushotzky}, R.~F. and {Szymkowiak}, A.~E.},
        title = "{Einstein Observatory Solid State Detector Observations of Cooling Flows in Clusters of Galaxies}",
    booktitle = {Cooling Flows in Clusters and Galaxies},
         year = 1988,
       editor = {{Fabian}, A.~C.},
       series = {NATO Advanced Study Institute (ASI) Series C},
       volume = {229},
        month = jan,
        pages = {53},
          doi = {10.1007/978-94-009-2953-1_6},
       adsurl = {https://ui.adsabs.harvard.edu/abs/1988ASIC..229...53M}
}

@ARTICLE{Suleimanov2019,
       author = {{Suleimanov}, Valery F. and {Doroshenko}, Victor and {Werner}, Klaus},
        title = "{Hard X-ray view on intermediate polars in the Gaia era}",
      journal = {\mnras},
         year = 2019,
        month = jan,
       volume = {482},
       number = {3},
        pages = {3622-3635},
          doi = {10.1093/mnras/sty2952},
archivePrefix = {arXiv},
       eprint = {1809.05740},
 primaryClass = {astro-ph.HE},
       adsurl = {https://ui.adsabs.harvard.edu/abs/2019MNRAS.482.3622S}
}

@ARTICLE{Luna2026YZRET,
       author = {{Luna}, G.~J.~M. and {Rawat}, N. and {Angeloni}, R. and {Orio}, M. and {Scaringi}, S. and {Dobrotka}, A. and {Magdolen}, J.},
        title = "{A very rapidly rotating white dwarf in nova YZ Reticuli}",
      journal = {arXiv e-prints},
         year = 2026,
        month = may,
          eid = {arXiv:2605.22802},
        pages = {arXiv:2605.22802},
          doi = {10.48550/arXiv.2605.22802},
archivePrefix = {arXiv},
       eprint = {2605.22802},
 primaryClass = {astro-ph.SR},
       adsurl = {https://ui.adsabs.harvard.edu/abs/2026arXiv260522802L}
}

@ARTICLE{Luna2026V1405,
       author = {{Luna}, G.~J.~M. and {Dobrotka}, A. and {Orio}, M.},
        title = "{A TESS view of post-eruption variability in novae V1405 Cas, V1716 Sco, and V1674 Her}",
      journal = {\aap},
         year = 2026,
        month = apr,
       volume = {708},
          eid = {A352},
        pages = {A352},
          doi = {10.1051/0004-6361/202557972},
archivePrefix = {arXiv},
       eprint = {2511.06399},
 primaryClass = {astro-ph.SR},
       adsurl = {https://ui.adsabs.harvard.edu/abs/2026A&A...708A.352L}
}

@ARTICLE{Imamura1983ApJ...268..291I,
       author = {{Imamura}, J.~N. and {Durisen}, R.~H.},
        title = "{X-ray spectra and light curves of accreting magnetic degenerate dwarfs}",
      journal = {\apj},
         year = 1983,
        month = may,
       volume = {268},
        pages = {291-308},
          doi = {10.1086/160954},
       adsurl = {https://ui.adsabs.harvard.edu/abs/1983ApJ...268..291I}
}

@ARTICLE{Protassov2002ApJ...571..545P,
       author = {{Protassov}, Rostislav and {van Dyk}, David A. and {Connors}, Alanna and {Kashyap}, Vinay L. and {Siemiginowska}, Aneta},
        title = "{Statistics, Handle with Care: Detecting Multiple Model Components with the Likelihood Ratio Test}",
      journal = {\apj},
         year = 2002,
        month = may,
       volume = {571},
       number = {1},
        pages = {545-559},
          doi = {10.1086/339856},
archivePrefix = {arXiv},
       eprint = {astro-ph/0201547},
 primaryClass = {astro-ph},
       adsurl = {https://ui.adsabs.harvard.edu/abs/2002ApJ...571..545P}
}
\bibliographystyle{aasjournal}



\end{document}